\documentclass[letter, longauth]{aa} 

\usepackage{natbib}
\newcommand{\Espresso}{\textsc{Espresso}}

\usepackage{graphicx}
\usepackage{hyperref}
\usepackage{txfonts}
\newcommand{\logg}{\ensuremath{\log g}}

\newcommand{\feh}{\rm [Fe/H]}
\newcommand{\teff}{T$_{\rm eff}$}

\definecolor{lightblue}{rgb}{.70,.95,1}

\begin{document} 

\title{The pristine nature of SMSS\,1605$-$1443 revealed by \Espresso \thanks{Based on ESPRESSO GTO collected under ESO programmes 1104.C-0350, 108.2268.001, P.I. P. Molaro. Based also on UVES data retrieved from the ESO archive under programme 105.20K7.001.}
}
   
\author{D.~S. Aguado\inst{1,2,3}, 
E. Caffau\inst{4}, 
P. Molaro\inst{5,6}, 
C. Allende Prieto\inst{3,7}, 
P. Bonifacio\inst{4}, 
J.~I. Gonz\'alez Hern\'andez\inst{3,7}, 
R. Rebolo\inst{3,7,8}, 
S. Salvadori \inst{1,2},
M. R. Zapatero Osorio\inst{9}, 
S. Cristiani\inst{5,6},
F. Pepe\inst{10},
N.~C. Santos\inst{11,12},
G. Cupani\inst{5,6},
P. Di Marcantonio\inst{5}
V. D'Odorico\inst{5,13,6},
C. Lovis\inst{10},
N.~J. Nunes\inst{14},
C. J. A. P. Martins\inst{11,15},
D. Milakovi\`c\inst{5,6,16},
J. Rodrigues\inst{11,12},
T.~M. Schmidt\inst{10,5},
A. Sozzetti\inst{17},
A. Su\'arez Mascare\~no\inst{3,7}
}


\institute{Dipartimento di Fisica e Astronomia, Universit \'a degli Studi di Firenze, Via G. Sansone 1, I-50019 Sesto Fiorentino, Italy.
\and 
INAF-Osservatorio Astrofisico di Arcetri, Largo E. Fermi 5, I-50125 Firenze, Italy.
\and
Instituto de Astrof\'{\i}sica de Canarias, V\'{\i}a L\'actea, 38205 La Laguna, Tenerife, Spain.
\and
GEPI, Observatoire de Paris, Université PSL, CNRS, 5 Place Jules Janssen, 92190 Meudon, France.
\and
INAF-Osservatorio Astronomico di Trieste, Via G.B. Tiepolo 11, I-34143 Trieste, Italy.
\and
Institute of Fundamental Physics of the Universe, Via Beirut 2, Miramare, Trieste, Italy.
\and
Universidad de La Laguna, Departamento de Astrof\'{\i}sica,  38206 La Laguna, Tenerife, Spain.
\and
Consejo Superior de Investigaciones Cient\'{\i}ficas, 28006 Madrid, Spain.
\and
Centro de Astrobiología (CSIC-INTA), Carretera Ajalvir km 4, 28850 Torrejón de Ardoz, Madrid, Spain.
\and
Département d’astronomie de l’Université de Genève, Chemin Pegasi 51, 1290 Versoix, Switzerland.
\and
Instituto de Astrof\'isica e Ci\^encias do Espa\c co, CAUP, Universidade do Porto, Rua das Estrelas, 4150-762, Porto, Portugal.
\and
Departamento de Física e Astronomia, Faculdade de Ciências, Universidade do Porto, Rua Campo Alegre, 4169-007 Porto, Portugal.     
\and
Scuola Normale Superiore P.zza dei Cavalieri, 7 I-56126 Pisa.
\and
Instituto de Astrofísica e Ciências do Espaço, Faculdade de Ciências da Universidade de Lisboa,
Campo Grande, PT1749-016 Lisboa, Portugal.
\and
Centro de Astrof\'{\i}sica da Universidade do Porto, Rua das Estrelas, 4150-762 Porto, Portugal 
\and
INFN, Sezione di Trieste, Via Valerio 2, I-34127 Trieste, Italy
\and
INAF - Osservatorio Astrofisico di Torino, Via Osservatorio 20, I-10025 Pino Torinese, Italy.
\\             
}   

\authorrunning{Aguado et al.\\}
\titlerunning{The pristine nature of SMSS\,1605$-$1443}

 
  \abstract
   {SMSS\,J160540.18$-$144323.1 is the carbon-enhanced metal-poor (CEMP) star with the lowest iron abundance ever measured, $\feh=-6.2$, which was first reported with the SkyMapper telescope. The carbon abundance is $\rm A(C)\approx6.1$ in the low-C band, as the majority of the stars in this metallicity range. Yet, constraining the isotopic ratio of key species, such as carbon, sheds light on the properties and origin of these elusive stars.}
   {We performed high-resolution observations of SMSS\,1605$-$1443 with the ESPRESSO spectrograph to look for variations in the radial velocity ($v_{rad}$) with time. These data have been combined with older MIKE and UVES archival observations to enlarge the temporal baseline. The $^{12}$C/$^{13}$C isotopic ratio is also studied to explore the possibility of mass transfer from a binary companion.}
   {A cross-correlation function against a natural template was applied to detect $v_{rad}$ variability and a spectral synthesis technique was used to derive $^{12}$C/$^{13}$C in the stellar atmosphere.}
   {We confirm previous indications of binarity in SMSS\,1605$-$1443 and measured a lower limit $^{12}$C/$^{13}$C$>60$ at more than a 3\,$\sigma$ confidence level, proving that this system is chemically unmixed and that no mass transfer from the unseen companion has happened so far.  
   Thus, we confirm the CEMP-no nature of SMSS\,1605$-$1443 and show that the pristine chemical composition of the cloud from which it formed is currently imprinted in its stellar atmosphere free of contamination.}
   {}

\keywords{stars: abundances – stars: Population II - stars: Population III – 
Galaxy: abundances – Galaxy: formation – Galaxy: halo
               }
               
\maketitle
%

\section{Introduction}\label{sec:intro}
Ultra, hyper, and mega metal-poor stars (corresponding to $\rm [Fe/H]<-4$,
$\rm[Fe/H]<-5$, and $\rm[Fe/H]<-6$, respectively) are messengers from the early Universe. They provide crucial clues as to the yields of the first stars, their initial mass function, and the first stages of their chemical enrichment. At the lowest iron abundances, we can find three groups: (a)\,stars with no carbon enhancement \citep[carbon-normal, see e.g.][]{caff11, sta18}; (b)\,stars enriched in carbon, $\rm [C/Fe]>+1$, and also enhanced in neutron-capture process elements $\rm [Eu/Fe]>+0.7$ and/or $\rm [Ba/Fe]>+1$, CEMP-$r$, and/or -$s$ \citep[see e.g.][]{hansen-r,han16I,holmbeck2020}; and (c)\,stars enriched in carbon $\rm [C/Fe]>+1$, but not in r- or s-process elements, the so-called CEMP-no group  \citep[see e.g.][]{spite13, yong13II, yoon16}. 
The first group, the carbon-normal metal-poor stars, is less populated towards lower metallicities  \citep[see e.g.][]{coh05, car12, pla14b} with important consequences on the likelihood of low-mass star-formation at earlier times \citep[see e.g.][]{brom03}.  The second group is made up  of CEMP-$s$ and CEMP-$r/s$ (also known as CEMP-$i$), and they are likely binary systems where carbon and  the heavy elements were produced by a more massive stellar companion and subsequently donated to the star \citep{her05, Jones2016}. This group is also characterised by absolute carbon abundances reaching values as high as the solar abundance (A(C)~$\sim 8.4$) \citep[][]{spite13, boni15, yoon16}. Finally, the CEMP-no group stars show lower carbon abundances (A(C)~$\sim 6.4$), with no evidence of over-abundances of neutron capture elements. They dominate the population at lower metallicities. When looked for, they were shown to be single and therefore it is not expected that this group is populated by binary systems that undergo mass transfer following the production of neutron capture elements \citep{ryan05,lucatello05,car08,zepeda2022}.

One remarkable exception is the well-known CEMP-no star HE\,0107$-$5240 with $\rm [Fe/H]=-5.4$, which was discovered by \citet{chris01b} and thought to be a single system for 20 years. However, after an exhaustive follow-up \citep{arentsen19, boni20, aguado22}, the binary nature of HE\,0107$-$5240 was revealed. \citet{aguado22} showed that the isotopic ratio $^{12}$C/$^{13}$C is high, indicating that HE\,0107$-$5240 did not undergo mass transfer.
According to the classification of \citet{bee05}, SMSS\,1605$-$1443 is a mega iron-poor star with $\rm [Fe/H]=-6.2\pm0.2$ and a carbon enrichment at $\rm [C/Fe]=+3.9\pm0.2$, as first reported by \citet{nordlander2019MNRAS.488L.109N}. At the time of writing, the available upper limits ($\rm [Sr/Fe]<+0.2$ and $\rm [Ba/Fe]<+1.0$) suggest this is a CEMP-no star. However, the ultimate nature of this star has been questioned by \citet{aguado22}, who found a significant $v_{rad}$ variation in the observations from Echelle Spectrograph for Rocky Exoplanets and Stable Spectroscopic Observations (ESPRESSO). Therefore, confirmation of the binarity of SMSS\,1605$-$1443 is required and further chemical analysis is necessary to address the  question of whether the chemical composition of SMSS\,1605$-$1443 is indeed pristine.
In this work, we integrate the kinematical analysis with archival UVES spectra and derive an informative lower limit of $^{12}$C/$^{13}$C. The relevance of these findings for understanding binarity among the most iron-poor stars known to date is discussed. 

\begin{figure}
\begin{center}
{\includegraphics[width=65 mm, angle=90,trim={ .cm .cm .3cm 0cm},clip]{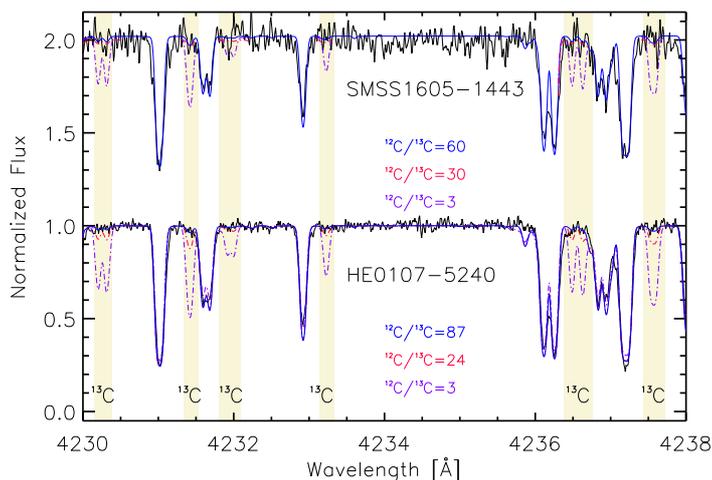}}
\end{center}
\caption{ Narrow region of the ESPRESSO combined spectrum of SMSS\,1605$-$1443 and HE\,0107$-$5240 around the G band. Three different synthetic models computed with {\tt SYNTHE} are also shown. }
\label{fig:carbon}
\end{figure}

\section{Observations and data reduction}\label{sec:observations}
To study and characterise SMSS\,1605$-$1443, we worked with three different sources of spectroscopic data: (a)\,intermediate spectral resolution ($R \approx 28,000$) observations with the  Magellan Inamori Kyocera Echelle (MIKE) spectrograph at the Magellan telescope  published by \citet{nordlander2019MNRAS.488L.109N}; (b)\,higher resolution ($R \approx 41,000$) follow-up with the Ultraviolet Visual Echelle Spectrograph (UVES) \citep{dek00} on the 8.2\,m Kueyen Very Large Telescope (VLT) publicly available from the European Southern Observatory (ESO) archive\footnote{http://archive.eso.org}; and (c)\,an ESO Guaranteed Time Observations (GTO) programme with the ultra-stable ESPRESSO spectrograph ($R \approx  138,000$) at VLT \citep{pepe21}. The MIKE data are not publicly available, and therefore we  refer to the original paper by \citet{nordlander2019MNRAS.488L.109N} and quote the results of the single observing night of September 1, 2018, summarised in Table \ref{table:rvs}.

Five additional UVES observations of SMSS\,1605$-$1443 taken between February 21, 2021 and September 6, 2021 are freely available. A $1\farcs0$ slit was used with $2\times2$ binning in dark sky conditions and a maximum airmass of $\sim1.4$. The setting used was dichroic,  $\#2$, with central wavelengths at 437\,nm for the blue arm and +760 for the red arm, providing a spectral coverage between 380 and 946\,nm. We corrected each spectrum for the barycentric velocity.  The average signal-to-noise per pixel in the spectra was $\sim$27 at 430\,nm and the seeing average of the five nights was $0\farcs82$ (see table \ref{table:rvs} for further details). The data were reduced using the REFLEX environment \citep{reflex} within the ESO Common Pipeline Library.

The  ESPRESSO observations were taken in the Single42 mode. ESPRESSO has  two fibres, one for the object and one for the  sky,  with a diameter of $140\,\mu m$ that corresponds to a 1\farcs0 aperture in the sky. The binning of the CCD was  $4\times2$ pixel and we selected the slow readout mode. The five spectra were taken in service mode with an individual exposure time of $3150$\,s, leading to data with S/N between 11 and 13 at 430\,nm. The observing criteria were:  seeing $\leq 1\farcs0$, airmass$\leq 1.5$, water vapour $\leq 30$\,mm, and a minimum lunar distance of $30^{\circ}$. The first observation (see table \ref{table:rvs}) was taken with higher seeing so that the exposure was repeated. 
Data reduction was performed by  using the automatic ESPRESSO pipeline, including sky subtraction, bias, and flat-fielding correction. The wavelength calibration combines a ThAr lamp  with a Fabry-P\'erot etalon \citep[][]{pepe13}. In Sec. \ref{ape:fig} we show the combined ESPRESSO spectrum.

\section{Analysis}\label{sec:analysis}
\subsection{Stellar parameters and radial velocity}

\citet{nordlander2019MNRAS.488L.109N} found good agreement between stellar parameters (effective temperature and surface gravity), which were derived both photometrically and spectroscopically, converging into \teff$=4850\pm100$\,K and \logg$=2.0\pm0.2$. Based on the \textit{Gaia} colours \citep{gaiaidr3} and applying the calibration by \citet{mucciarelli21}, we derived \teff$=4830\pm100$\,K. Additionally, based on the \textit{Gaia} parallax (with a zero point), we found \logg$=1.8\pm0.2$. Both parameters are in excellent agreement with the original ones; therefore, we assumed those from \citet{nordlander2019MNRAS.488L.109N}. We marginally detected the strong iron line at 404.5\,nm, in contrast with the ten \ion{Fe}{i} lines detected in the blue region of the MIKE spectrum. This is not surprising since ESPRESSO is less efficient in the blue, where the strongest \ion{Fe}{i} are. Therefore, we also adopted the metallicity from \citet{nordlander2019MNRAS.488L.109N}, $\feh=-6.2$.

To derive radial velocities from UVES and ESPRESSO, we followed the same iterative process explained in \citet{aguado22}. We used a template based on the same set of stellar parameters and performed a cross-correlation function (CCF) against each individual exposure in the range 420$-$431\,nm. The CCF was done in Fourier space using the algorithm from \citet{tonry79} implemented in the {\tt fxcor} package within {\tt IRAF} \citep{tod93}. Then, by correcting all the spectra and by combining them, we created a natural template in the rest of the frame for both instruments. Finally, we cross-correlated these templates with all individual spectra and derived final $v_{rad}$ measurements. This step allowed us to significantly reduce the derived uncertainties that are summarised in Table \ref{table:rvs}. While with UVES we get uncertainties of a few hundred of metres per second, with the ESPRESSO data these are an order of magnitude lower (tens of metres per second). 
On the other hand, as previously explained, the $v_{rad}$ measurement from MIKE data is directly taken from \citet{nordlander2019MNRAS.488L.109N} (See table \ref{table:rvs}). Although no error bar is provided, from our experience with MIKE with giant metal-poor stars, we estimate it as 1\,km$\,$s$^{-1}$.
Finally, we have searched the literature for further $v_{rad}$ measurements associated with this object. Unfortunately, SMSS\,1605$-$1443 is not bright enough to be observed with the Radial Velocity Spectrometer \citep[RVS, ][]{gaia_rvs} within the {\it Gaia} mission, and no data were found. We also checked the Survey of Surveys \citep[SoS,][]{Tsantaki22} including $v_{rad}$ measurements from different spectroscopic surveys, and no measurements were available.

\subsection{Carbon isotopic ratio $^{12}$C/$^{13}$C}\label{caff}

Metallic absorptions in the ESPRESSO spectra are almost exclusively due to carbon (CH) features. The extremely low metallicity of this star and the very high carbon enrichment account for this fact. However, the resolution and the relatively high quality of the ESPRESSO spectrum allowed us to derive the absolute carbon abundance A(C) and set a critical lower limit on the carbon isotopic ratio $^{12}$C/$^{13}$C. By assuming stellar parameters (\teff, \logg, and \feh) discussed in Sec. \ref{sec:analysis}, microturbulence of 1.8\,km$\,$s$^{-1}$, and a [$\rm \alpha/Fe]=+0.4$, we computed a stellar model with the {\tt ATLAS 9} model atmosphere code using an opacity distribution function of metallicity $-$5.5\footnote{\url{https://wwwuser.oats.inaf.it/castelli/odfnew.html}}. Using this model atmosphere, we computed a grid of synthetic spectra using the with {\tt SYNTHE} code \citep{kur05, sbo05}. To derive the absolute carbon abundance, we fitted the ESPRESSO spectrum with the collection of synthetic spectra in the G band, between 420 and 427\,nm, by minimisation of the $\chi^{2}$. The best fit provides $\rm A(C)=5.97\pm 0.10$. If we assume a solar carbon abundance A(C)$_{\odot}=8.43$ from \citet{asplund09},  we end up with $\rm [C/Fe]=+3.75\pm 0.20$, which is in good agreement with the value originally reported of $\rm [C/Fe]=+3.9\pm 0.2$. \citet{nordlander2019MNRAS.488L.109N} refrained from giving a lower limit for the isotopic ratio $^{12}$C/$^{13}$C due to the resolution of the MIKE spectrum. However, with the ESPRESSO data, we are able to give an informative value to $^{12}$C/$^{13}$C. In Fig. \ref{fig:carbon} we show the spectral region within the G band where some of the most prominent $^{13}$C features are located. However, to optimise the calculation of a lower limit, we decided not use individual lines but the entire information on $^{13}$C present in the range 415$-$435\,nm with a Markov chain Monte Carlo (MCMC) self-adaptative algorithm \citep{vru09} available within the {\tt FERRE} code \citep{alle06}. We conclude that  $^{12}$C/$^{13}$C $> 60$ is a robust lower limit with a statistical significance larger than $3\sigma$. Other values ($^{12}$C/$^{13}$C=3 or 30) with higher $^{13}$C abundance have been ruled out (red and purple lines, respectively). The complete MCMC methodology already tested by \citet{agu19a} is detailed in Appendix \ref{ape:mcmc}.

\begin{figure}
\begin{center}
{\includegraphics[width=65 mm, angle=90]{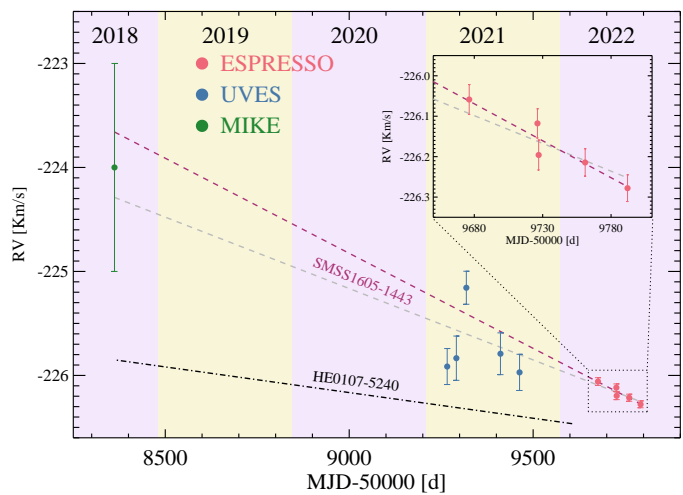}}
\end{center}
\caption{ Radial velocity points of SMSS\,1605$-$1443 versus modified Julian date (MJD), obtained with MIKE, UVES, and ESPRESSO. Grey and purple lines are a linear fit considering all instruments and ESPRESSO alone, respectively. The observed trend in HE\,0107$-$5240 with ESPRESSO is also shown (black line, which has been vertically shifted to improve visibility). }
\label{fig:vrad}
\end{figure}

\begin{figure}
\begin{center}
{\includegraphics[width=65 mm, angle=90,trim={ .cm 0.cm 0.0cm 0.2cm},clip]{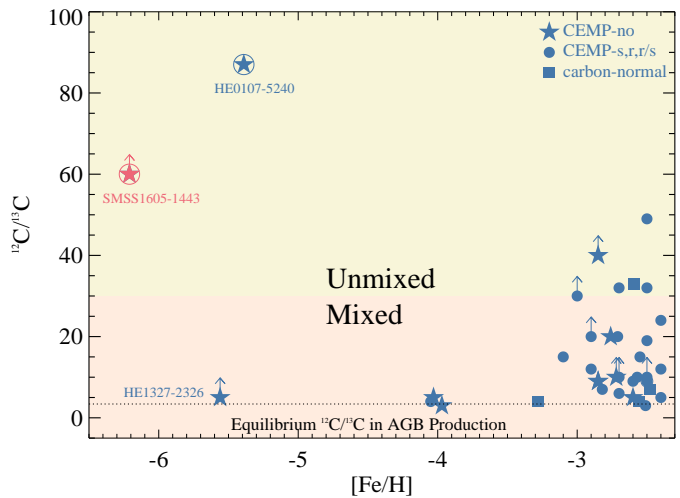}}
\end{center}
\caption{ $^{12}$C/$^{13}$C$-$[Fe/H] plane for SMSS\,1605$-$1443 (red symbols) and other stars from the literature (blue symbols). According to \citet{spite06}, the $^{12}$C/$^{13}$C$=30$ value is also shown.}
\label{fig:iso}
\end{figure}

\section{Discussion and conclusion}\label{sec:discussion}

\subsection{The $v_{rad}$ variability of SMSS\,1605$-$1443}
The re-normalised unit weight error (RUWE) published by \textit{Gaia} EDR3 \citep{gaiaidr3} is 1.06 and does not allow us to conclude about binarity whatsoever. This is not surprising since the star is at a distance of $\sim8.9$\,kpc \citep{bay18}, which is far from the sensitivity range of \textit{Gaia} parallaxes. However, an indication of radial velocity variability was reported by \citet{aguado22} based on the same five ESPRESSO observations. Thanks to the ultra stability of this spectrograph, we can see the decreasing trend shown in Fig. \ref{fig:vrad} (red points) and this would be sufficient to claim a $v_{rad}$ variation. However, a longer time range would be highly desirable to confirm this hypothesis and to have clues as to the period. Luckily, the early MIKE data (green point in Fig. \ref{fig:vrad}) and the subsequent UVES observations (blue points) all together span five years of radial velocity data. The error bars corresponding to different instruments are significantly different. MIKE's are about ten times larger than UVES's, and UVES's are ten times larger than ESPRESSO's. However, all of them are compatible with the same decreasing trend represented by the pink dashed line in Fig. \ref{fig:vrad}. In other words, comparing the slopes of the $v_{rad}$ variation that we see when we consider only the ESPRESSO points ($\sim1.8\pm0.4\,\rm m\,s^{-1}\,$d$^{-1}$) is compatible within the errors with the one calculated considering the measurements from the three instruments ($\sim1.4\pm0.2\,\rm m\,s^{-1}\,$d$^{-1}$). Therefore, the additional data we provide in this analysis confirm the previous claim by \citet{aguado22}, and SMSS\,1605$-$1443 unambiguously shows a decreasing pattern in $v_{rad}$. The underlying assumption of linear variation is, in general, not realistic, and it accounts for the small difference between the two calculated ratios.

The time baseline covered by the $v_{rad}$ observations is too short for a reliable Keplerian analysis of the SMSS\,1604$-$1443 system. 
We ran a preliminary set of orbital fit simulations on the joined radial velocity data (including MIKE, UVES, and ESPRESSO) within a Bayesian scheme with wide priors for all binary parameters, including the orbital period, semi-amplitude velocity, and eccentricity. The result points to long orbital periods above 8~yr, as well as low eccentricity and semi-amplitude velocity. New ESPRESSO observations are required during the next years to better constrain the orbital period and other binary parameters.

\subsection{The pristine nature of  SMSS\,1605$-$1443}
In their discovery paper, \citet{nordlander2019MNRAS.488L.109N} provided informative upper limits for n-capture elements, $\rm [Sr/Fe]<+0.2$ and $\rm [Ba/Fe]<+1.0$, suggesting this star belongs to the CEMP-no group. However, being a binary system, some mass transfer from a  companion could have happened. If this were the case, the chemical composition of  SMSS\,1605$-$1443 would not reflect that of the natal cloud from which it was formed. The carbon isotopic ratio $^{12}$C/$^{13}$C is frequently employed as an asymptotic giant branch (AGB) donation tracer since large amounts of the originally produced $^{12}$C are converted into  $^{13}$C thanks to the CN cycle \citep{wannier80} during the red giant branch (RGB) \citep{weiss2000}. Consequently, the isotopic ratio in evolved stars rapidly decreases to low values, typically $^{12}$C/$^{13}$C$\sim3-4$. Therefore, if SMSS\,1605$-$1443 underwent mass transfer from a former giant companion, we should see low values of $^{12}$C/$^{13}$C within its stellar atmosphere. However, as shown in Sec. \ref{sec:analysis}, the obtained lower limit, $^{12}$C/$^{13}$C>60, is much higher than the equilibrium relation mentioned.

Old, metal-poor stars with no significant contribution from CN-processed material are considered chemical unmixed \citep{spi05}. Therefore, SMSS\,1605$-$1443, with an amount of iron 1.5 million times lower than the Sun, is a real second-generation star exclusively polluted by previous SN events, though living with an unseeing companion. 
In Fig. \ref{fig:iso} we show the existing metal-poor stars with $^{12}$C/$^{13}$C measurements or lower limits \citep{ kipper1994,Pilachowski1997, hill02, fre06, Sivarani2006, behara2010,Masseron2012,pla15, han16I,spite21, aguado22}. The number of points is still small due to the difficulty of measuring $^{12}$C/$^{13}$C, but the data suggest that a more metal-poor unmixed star tends to form more in binary systems than other stars born in relatively more metal-rich regimes \citep{badenes18, arentsen19}. 

It is rather difficult to place constraints on the binary mass function of the binary system $f({\rm M})$ due to the relatively short time range of available $v_{rad}$ measurements. However, we can discuss possible scenarios. There are several mechanisms of mass transfer in binaries depending on the relative distance scales of the system.  Regardless of whether the binary separation ($\rm D_{b}$) is slightly larger than the dust formation radius ($\rm R_{dust}$), the accretion is in the form of wind Roche lobe overflow. However, if $\rm D_{b}<\rm R_{dust}$, Roche lobe overflow may happen. Finally, if $\rm D_{b}>>\rm R_{dust}$, the accretion on to the secondary is a driven Bondi–Hoyle mechanism \citep[see e.g.][]{yuhan2002,chen2017}. Therefore, even in wide binaries with $\rm D_{b}$ of the order of tens of AU, stars could eventually receive mass from an AGB companion.  Since such a possibility is discarded on the basis of its chemical composition, we conclude that the unseen component of the SMSS\,1605$-$1443 system has never reached the giant phase. Our favoured explanation is that the unseen companion is a low-mass star that has not left the main sequence yet. We estimate that the mass of such a companion could not have exceeded 0.8\,$\rm M_{\odot}$; otherwise, the companion would have plenty of time to start the H-shell burning phase and quickly rise up to the giant branch.
We find clear similarities between SMSS\,1605$-$1443 and HE\,0107$-$5240, another binary system likely composed of two low-mass stars. Unfortunately, due to the faint magnitude of SMSS\,1605$-$1443 (G$_{Gaia}=15.4$), no spectral distribution of energy could be analysed since no GALEX ultraviolet data \citep{galex05} are available for this star. Therefore, confirmation of the precise nature of its companion could not be obtained. More observations with ultra-stable ESPRESSO spectroscopy are required in a longer time interval to determine an orbit for the system.

\subsection{The binary fraction in CEMP-no stars}

Different origins have been proposed to explain the high carbon enrichment in extremely metal-poor stars. In particular, for CEMP-no stars, it was widely believed that the majority of them are not part of binary systems \citep{ryan05, starkenburg14, han16I}. However, recent studies are now challenging this picture, at least at the metal-poor end of the Galactic halo \citep{arentsen19, boni20, aguado22}. The fact that out of eight stars with $\rm [Fe/H]<-4.5$ already observed with ESPRESSO, two of them are binaries increases the expected ratio of this rare class. We also notice that binarity could not be excluded from the other six stars with the data in hand \citep{aguado22}. Evidence of binarity among the most iron-poor CEMP-no stars is potentially very interesting. Indeed, theoretical models that study the fragmentation properties of star-forming clouds in the presence of dust \citep[e.g.][]{Omukai2005} found that gaseous environments with 10$^{-5}$ Z$_\odot$ < Z < 10$^{-2}$ Z$_\odot$ can fragment into very small sub-solar clumps, while the mass of these clumps increases for Z > 10$^{-2}$ Z$_\odot$. This implies that many low-mass stars can form in very metal-poor environments and thus that binary systems might be more common among very metal-poor stars \citep[][]{Matsukoba2022}.

\subsection{On the origin of carbon}

\citet{spi13} noted that carbon  is not following the general iron decrease, but it remains rather constant at the lowest metallicities. They suggested the existence of two  different carbon levels: one with relatively high carbon abundance, $\rm [C/H] \approx -2.0$, and a second with about $\rm [C/H] \approx -3.5$. Carbon has a different origin  in the two bands: in the high-C band, it is acquired from an AGB companion, together with n-capture elements. In the low-C band, it is instead representative of the environment of formation. However, what the stellar objects are that polluted this birth environment is still unclear. 

Low-energy primordial faint supernovae (SNe) with mixing and fallback have been the first sources proposed to enrich the environment of formation of CEMP-no stars \citep[e.g.][]{ume03,umeda2005,iwamoto2005,tominaga2007,ishi14,boni15}.
Recent models comparing the chemical abundances of five CEMP-no stars and the predicted SN yields confirm that SNe with a stellar mass of 11-22\,M$_{\odot}$ and low explosion energies, 0.3-1.8$\times$10$^{51}$\,erg, are possible progenitors of CEMP-no stars \citep{Almusleh2021AN....342..625A}. However, massive and fast-rotating low-metallicity {`spinstars'} have also been proposed as possible sources of enrichment for the birth environment of CEMP-no stars \citep{Meynet2006A,maeder2015A&A...576A..56M}. Ultimately, both solutions are plausible, and cosmological models can indeed successfully reproduce the observed fraction of CEMP-no stars at different [Fe/H] by including either primordial faint SNe \citep[e.g.][]{Bennassuti2017} or spinstars \citep[e.g.][]{Liu2021}. 
However, it has been shown that spinstars could produce a large amount of $^{13}$C \citep{Meynet2006A,Limongi2018,maeder2015A&A...576A..56M}, while zero-metallicity (faint) SNe do not \citep{HegerWoosley2010}.

Hence, our new results can allow us to finally constrain the progenitors of CEMP-no stars. SMSS\,1605$-$1443 shows $^{12}$C/$^{13}$C>60 and in the CEMP-no star HE\,0107$-$5240 this ratio is $87\pm6$ \citep{aguado22}. These high values in the carbon isotopic ratios rule out the significant production of $^{13}$C. Thus, the first generation of stars that polluted the gas from which CEMP-no stars are born does not seem to have made any significant amount of $^{13}$C. For this reason, the faint SNe hypothesis is favoured against spinstars. We note that \cite{hansen2015ApJ...807..173H,norris2013ApJ...762...28N} measured low $^{12}$C/$^{13}$C in a sample of CEMP-no stars. However, these stars also show [C/N] $<$ 0, which is evidence of internal mixing according to \citet{spite06}, and thus of a conversion of $^{12}$C into $^{13}$C (see also our Fig. \ref{fig:iso}).

The old, mega metal-poor star SMSS\,1605$-$1443 ($\rm [Fe/H]=-6.2$) is a unique single-lined, chemically unmixed binary star. Therefore, its chemical signature is likely reflecting the composition of the molecular natal cloud. New high-resolution ESPRESSO observations during the next years are required to constrain the orbital parameters.
Finally, the new generation of not only spectroscopic but also photometric surveys such as {\it Pristine} \citep{sta17I}, {\it S}$-$Plus \citep{splus2019}, and {\it J}$-$Plus \citep{cenarro19} will help to identify new CEMP-no candidates at $\rm [Fe/H]<-4$ and shed light on the origin of these elusive fossils.



\begin{acknowledgements}
DA ans SS acknowledge support from the ERC Starting Grant NEFERTITI H2020/808240.
EC acknowledges support from the French National Research Agency (ANR) funded project ``Pristine'' (ANR-18-CE31-0017).
JIGH, CAP, ASM and RR acknowledge financial support from the Spanish Ministry of Science and Innovation (MICINN) project PID2020-117493GB-I00.
MRZO acknowledges financial support from the Spanish Ministry of Science and Innovation through project PID2019-109522GB-C51.
ASM acknowledges financial support from the Spanish Ministry of Science and Innovation (MICINN) under 2018 Juan de la Cierva program IJC2018-035229-I. ASM acknowledge financial support from the Government of the Canary Islands project ProID2020010129.
FPE, CLO, and TMS would like to acknowledge the Swiss National Science Foundation (SNSF) for supporting research with ESPRESSO through the SNSF grants nr. 140649, 152721, 166227, 184618, and 193689. The ESPRESSO Instrument Project was partially funded through SNSF’s FLARE Programme for large infrastructures.
DM is also supported by the INFN PD51 INDARK grant.
This work was financed by FCT - Fundação para a Ciência e a Tecnologia under projects UIDB/04434/2020 \& UIDP/04434/2020, CERN/FIS-PAR/0037/2019, PTDC/FIS-AST/0054/2021.
CJM also acknowledges FCT and POCH/FSE (EC) support through Investigador FCT Contract 2021.01214.CEECIND/CP1658/CT0001.

\end{acknowledgements}
%
%

\bibliographystyle{aa}
\bibliography{biblio}

\begin{appendix}
\section{Co-added ESPRESSO spectrum and the table with $v_{rad}$ measurements.}\label{ape:fig}  
The combined ESPRESSO spectrum of  SMSS\,1605$-$1443 is shown together with a table with all $v_{rad}$ measurements described in Sec \ref{sec:analysis}.
\begin{figure*}
\begin{center}
{\includegraphics[width=130 mm, angle=90]{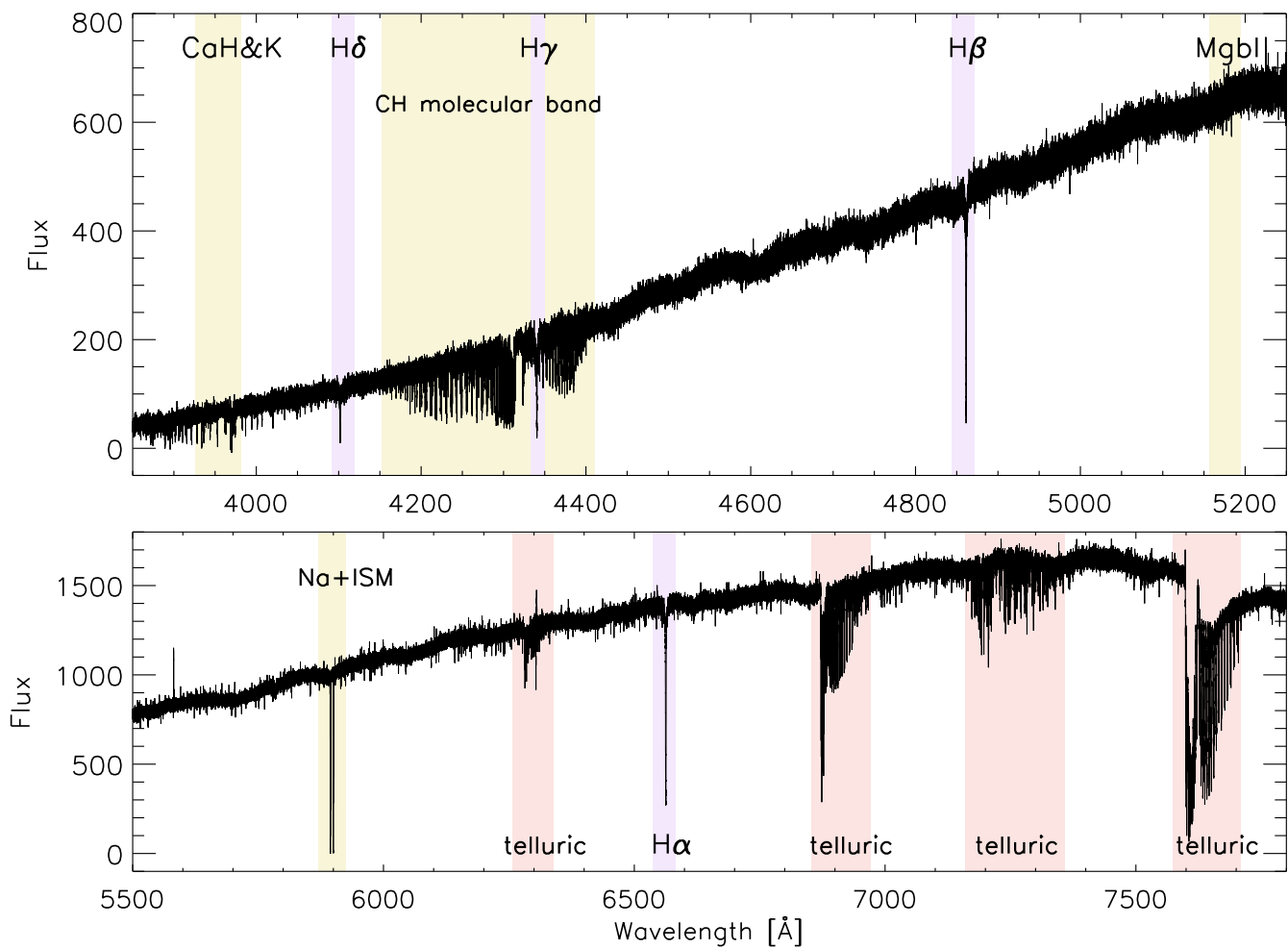}}
\end{center}
\caption{Co-added ESPRESSO spectrum of SMSS\,1605$-$1443 (black  lines). Absorptions corresponding to the Balmer series and \ion{Ca}{ii} H\&K region are marked in red, purple, and yellow, respectively. Other metallic absorptions and the G band are also labelled. Regions with strong telluric contamination are masked in red.}
\label{comp}
\end{figure*}

\begin{table*}
\begin{center}
    
\caption{\label{table:rvs} Radial velocities measurements from ESPRESSO, UVES and Mike.}
\scriptsize
\hspace*{0cm}\resizebox{1.\linewidth}{!}{
\begin{tabular}{lcccccccccc}
 \hline
 Instrument & $v_{rad}$ & error & MJD\tablefootmark{b} & MODE&$ \lambda/\delta\lambda$&t$_{exp}$&S/N&binning&seeing&Comment\\
   & (km$\,$s$^{-1}$) & (km$\,$s$^{-1}$) & $-50000$ && &s &&&arcsec& \\
 \hline
  \hline
Mike&$-$224      & 1\tablefootmark{a}& 8362.0  & AB    &28,000 & 1800& 10&   2$\times$2& --&\citet{nordlander2019MNRAS.488L.109N}        \\ 
UVES&$-$225.914 & 0.173& 9266.351&DIC\#2& 41,000& 2990& 27&   2$\times$2& 0.75               &\\    
UVES&$-$225.834 & 0.213& 9291.259&DIC\#2& 41,000& 2990& 24&   2$\times$2& 0.91                &  \\  
UVES&$-$225.157 & 0.159& 9318.365&DIC\#2& 41,000& 2990& 30&   2$\times$2& 0.77               &\\    
UVES&$-$225.791 & 0.201& 9411.142&DIC\#2& 41,000& 2990& 27&   2$\times$2& 0.91             &\\    
UVES&$-$225.971 & 0.174& 9463.006&DIC\#2& 41,000& 2990& 28&   2$\times$2& 0.77                &\\    
ESPRESSO&$-$226.059  &  0.037 & 9676.277 & HR42&145,000 &3150&11 &4$\times$2& 1.23  &\\    
ESPRESSO&$-$226.118  &  0.036 & 9726.189 & HR42&145,000 &3150&12 &4$\times$2& 0.68  &  \\  
ESPRESSO&$-$226.196  &  0.037 & 9727.060 & HR42&145,000 &3150&13 &4$\times$2& 0.92  &\\    
ESPRESSO&$-$226.214  &  0.030 & 9761.059 & HR42&145,000 &3150&13 &4$\times$2& 0.97  &\\    
ESPRESSO&$-$226.278  &  0.033 & 9792.055 & HR42&145,000 &3150&13 &4$\times$2& 0.79  &\\    
\hline
\end{tabular}}
\tablefoot{\tablefoottext{a}{Uncertainty no originally reported but assumed for the purpose of this work}\tablefoottext{b}{Modified Julian date at the start of observation.}}
\end{center}
\end{table*}

\section{$^{12}$C/$^{13}$C calculation with an MCMC methodology}\label{ape:mcmc}
To derive a constringent $^{12}$C/$^{13}$C lower limit, we employed an MCMC methodology already described in \citet{agu19a}. First, following the recipe presented in Sec. \ref{caff}, we computed a set of synthetic models with different values of $^{12}$C/$^{13}$C, from ten to 90 in steps of ten. The rest of the stellar parameters, effective temperature, surface gravity, metallicity, and absolute carbon abundance  were fixed to the values from Sec. \ref{sec:analysis}. Then we packaged the stellar models in a {\tt FERRE}\footnote{{\tt FERRE} is available from \url{http://github.com/callendeprieto/ferre}} readable shape. 

\begin{figure}
\begin{center}
{\includegraphics[width=0.75\hsize, angle=90]{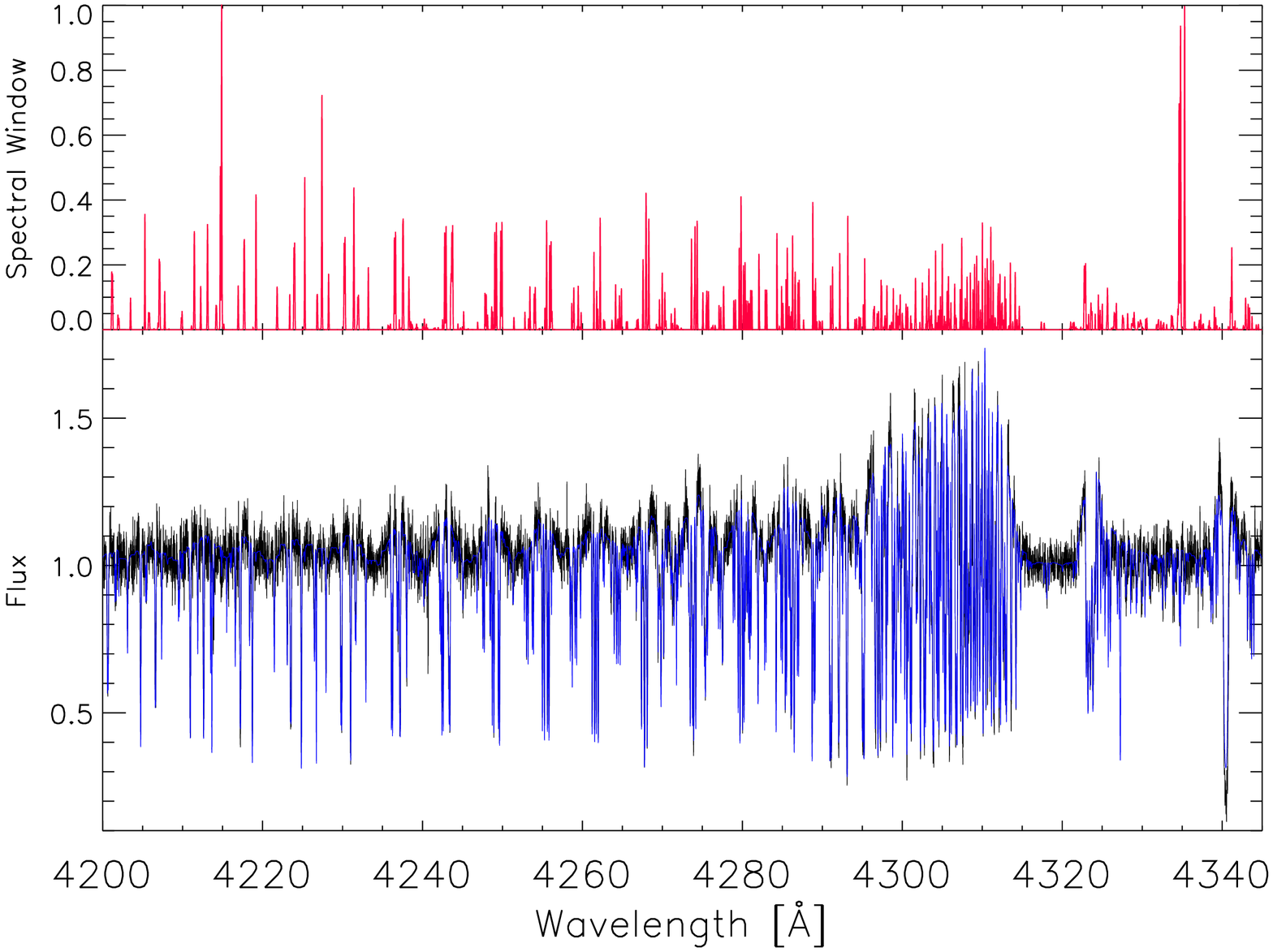}}
\end{center}
\caption{Upper-panel: Weights employed in the  {\tt FERRE} analysis (red). Lower-panel: Combined ESPRESSO spectrum around the G band (black) and the best fit derived with {\tt FERRE} (blue). Both the data and the best fit were normalised with a running mean filter with a 300\,pixel window.}
\label{fig:fit}
\end{figure}

\begin{figure}
\begin{center}
{\includegraphics[width=0.75\hsize, angle=90]{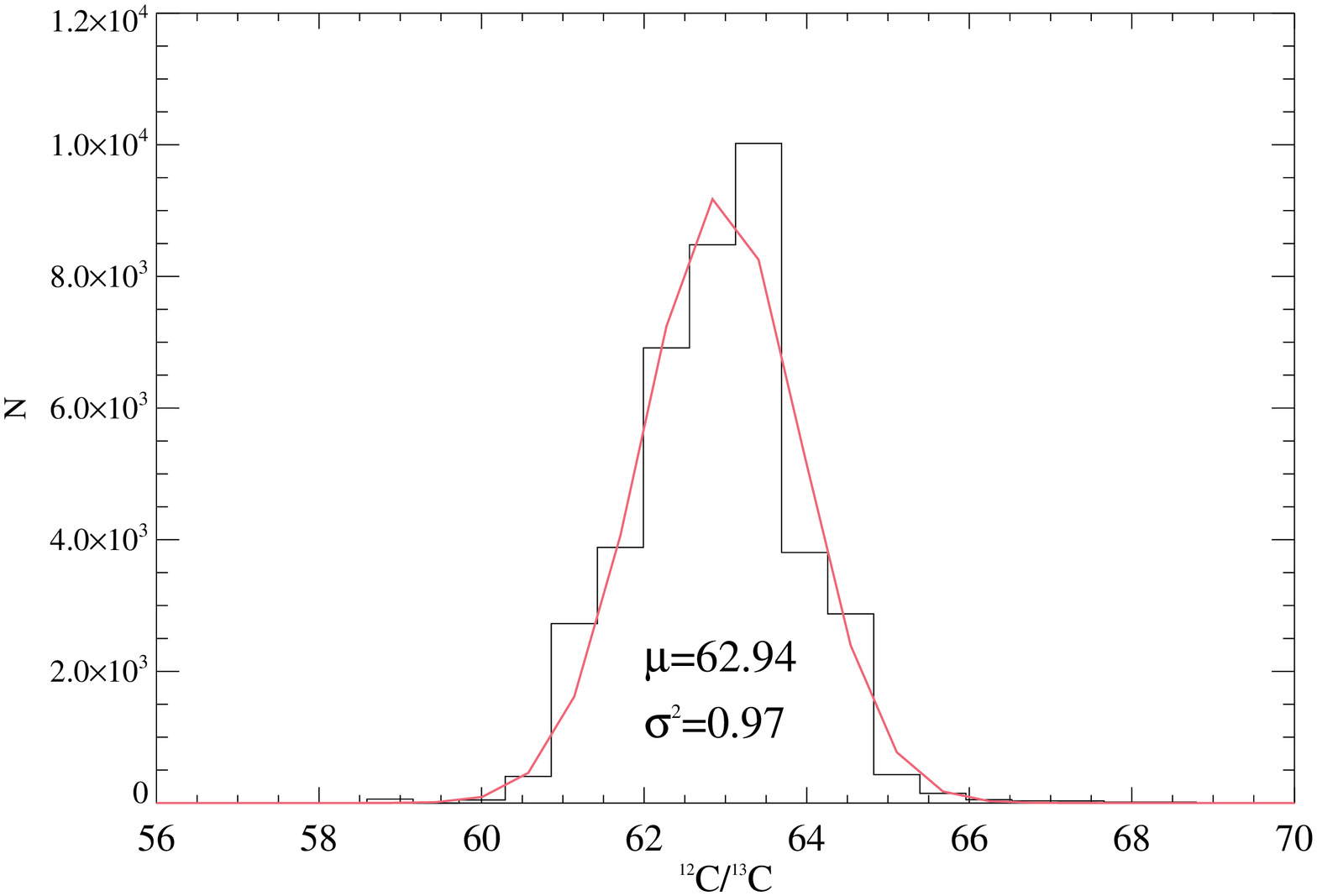}}
\end{center}
\caption{Sample distribution of Markov chain Monte Carlo experiments vs  $^{12}$C/$^{13}$C computed with {\tt FERRE} (black histogram) and the modelled Gaussian distribution with parameters (red line).}
\label{fig:mcmc}
\end{figure}

Then, to identify the regions within the G band more sensible to $^{12}$C/$^{13}$C variations, we subtracted the   $^{12}$C/$^{13}$C=90 normalised model from the $^{12}$C/$^{13}$C=10 model. The result was scaled between 0 and 1 (top panel of Fig. \ref{fig:fit}, in red). This difference was used to weigh the $\chi$-squared evaluation: more weight was given to frequencies that contain more information on the $^{12}$C/$^{13}$C ratio. {\tt FERRE} is equipped with an option to read these weights and use them.

Subsequently, we normalised both the set of models and the ESPRESSO data by using a running mean filter with a 300\,pixel window and indicated the code to fit the data with an MCMC self-adaptative randomised subspace sampling algorithm \citep{vru09}. Ten chains of 5,000 experiments each (discarding the first 500) were run following this methodology. The data and the derived best fit is shown in Fig. \ref{fig:fit}, in the lower panel. In Fig. \ref{fig:mcmc} we show the sample distribution of the MCMC experiments versus the most likely  $^{12}$C/$^{13}$C value. The distribution is somewhat skewed towards the higher values, and they could be approximated by a Gaussian with a central value of 62.94 and a variance of 0.97.

Although the result seems to be very robust with this methodology, we note that the spectral sensitivity at such high values of $^{12}$C/$^{13}$C (i.e. low values of $^{13}$C) is relatively low. Thus, we worry about possible uncontrolled systematics and prefer to consider it as a lower limit.
Then, considering the statistical parameters $\mu$ and $\sigma^{2}$, we derived  $^{12}$C/$^{13}$C>60 at more than a 3\,$\sigma$ confidence level. Finally, we considered the central value of the distribution as a tentative detection of $^{13}$C with $^{12}$C/$^{13}$C$=62.94$ and $\sigma^{2}=0.97$.
\end{appendix}

\end{document}